\documentstyle[12pt]{article}
\newcommand{\be}{\begin{equation}}
\newcommand{\ee}{\end{equation}}
\newcommand{\email}{e-mail:~serg@ptslab.ioffe.rssi.ru}
\newcommand{\SR}{{\it Sankt--Petersburg, Russia}}
\author{Serge B. Afanas'ev\\ \SR\\ \email}

\title{The Nambu--Goto action as the one of the
quantized space--time excitation.}

\begin{document}
\maketitle
\vskip10mm
\begin{abstract}
The concept of the quantized
space--time of the formless finite fundamental elements
is suggested. This space--time can be defined as a set of continual
space--time coverings by simply connected non--overlapping regions of
any form and arbitrary sizes with some probability measure. The functional
integrating method and the space--time action problem are analyzed.
A string in this space--time is considered as an excitation of a number of
fundamental elements forming one--dimensional curve.
It is possible to direct the way for the volume
term of the space--time
action to yield the Nambu--Goto action with this consideration.
\end{abstract}

\vskip10mm
Classical space--time is the continuum which is an
arena for particles motion at their interaction. This
concept of the space--time description remains in the quantum
mechanics and the quantum field theory, where the
space--time properties do not depend on particles
structure and propagation, but particles are described
by wave functions that have finite values in all
space--time. Connection of the space--time structure
and the matter is shown in the theory of gravitation.
Classical gravitation theory space--time is the
continuum also, but the quantum description of
space--time properties at small distances is
connected with the fundamental length ($l_f$)
existence and
discrete structure of the space--time.

The lattice space--time with the fixed lattice is most
commonly investigated~\cite{lat}.  But the space--time with
the fixed lattice consideration leads to several
problems. The first and seemingly the most essential
problem is the passage from the lattice space--time to the
continuum in the limit $l_f \to 0$. The lattice
space--time has the power of a countable set.  Any
subdivision of a lattice yields a set with the same
power.  Thus in the limit $l_f \to 0$ any lattice
space--time with any subdivision remains a set with the
power of a countable set.  The second, the resulting
equations in the lattice space and other spaces with
determined fundamental element form depend on form of
a fundamental element.  The third, the equations in the
lattice space are non-invariant under the continual
symmetry operations.

Discrete geometry has been developed in the direction
of the formless fundamental element in the last thirty
years.  The Regge calculus~\cite{regge} and the
space--time foam idea~\cite{hawking,foam} made the first steps on
this way.  In refs.~\cite{yak} the stereohedra
space is investigated. This space fundamental element
has some set of forms. Random lattice field theory is
analyzed in~\cite{chiu}.  Quantum configuration space
investigated in~\cite{asht} is the method of quantized
space--time description based on the not--fixed
"floating lattice".

The alternative approach to the particle physics
geometrical base is the superstring theory~\cite{gsw}.
Strings are fundamental objects, their excited
states describe all particles. But the strings
are described as the geometrical objects that
propagate in the background continual space--time.
Therefore strings can be considered as the basic
objects for fundamental particles description, but
they cannot be considered as the fundamental units of
the space--time.

In this work the concept of the space (space--time)
that consists of
the formless finite fundamental elements (FFFE) is
suggested.  On the one hand it makes the base for
space--time quantization. Consideration of this
concept leads to the idea of the space--time action with
one of the terms proportional to the volume of
FFFE. This method of space--time quantization is
a consistently geometrical approach to the physics of
fundamental particles and interactions. In this
space--time the particles are excited states of the
fundamental elements, and the interactions are
connected with the transformations of the Riemannian
space--time of FFFEs, that describes the space--time
with particle--like excitations.

On the other hand the strings in this space--time
picture can be considered as propagating excitations
of any number of FFFEs, forming one- dimensional
space--like curve (in the sense of FFFE space--time).
The volume term of the space--time action of this
excitation yields the Nambu--Goto action term.
This picture of the quantized
space--time and the strings as this space--time
excitations enables to connect these approaches and
assigns the meaning of fundamental space--time units
for the strings.

The space of FFFEs can be defined as the set of
coverings of the continual space by any number of
non-overlapping simply connected regions of any form
and arbitrary sizes. This set is provided with the
probability measure, i.e. each covering contributes to
the space with some probability.
This measure enables
the calculations based on this coverings set.
Obviously, the probability measure like that
the
average values of sizes of FFFEs are equal to $l_f$,
and the average number of FFFEs localized in the
continual space region by the volume $V$ is $N = [V
(l_f^n)^{-1}]$.  But the configurations with
the FFFEs sizes greatly
different from $l_f$ also have the finite
probabilities. For example, the configurations where
one fundamental element expands on all the space--time (or
investigated manifold),
or the configuration of continual
space--time region itself, i.e. covering this region
by points.  This set of coverings have the power of
continuum.  Therefore limit passage from the space of
FFFEs to the continual space can be carried out
correctly.

The general construction for calculations on this
coverings set is the continual (functional) integral.
In agreement with the central idea of the continual
integral theory the calculation of quantum quantities
is the integrating over all possible configurations of
the space--time of FFFEs (i.e. coverings of the
space--time) with  the corresponding
probability measure taken into account.

Consider the general construction of a functional
integral. In the plane space--time it is:
\be
Z = \int {\cal D} V e^{-S(s_i)},
\ee
where ${\cal D} V$ is a measure on the set of
coverings, $S(s_i)$ is the plane space--time vacuum
action, $s_i$ is the set of the element parameters (sizes,
areas, volume). Here integrating is over all coverings
of the continual space by non-overlapping simply
connected regions of any forms and any sizes. The average
value of a function on a separate FFFE is defined by
\be
\label{func}
<f({\{a\}})> = {{\int {\cal D} V
e^{-S(s_i)} f_{\{a\}}(x^i)} \over {\int {\cal D} V
e^{-S(s_i)}}},
\ee
where $f_{\{a\}}(x^i)$ are values of
the function $f$ at regions of coverings set which
forms the element with FFFE space--time coordinates
$\{a\}$, $f(x^i)$ is a function defined in the
continual space.

The functional integral construction requires the
information about the action.  The space--time vacuum
action of the Minkowski space--time
can depend on fundamental elements
geometrical characteristics
only: volume,
$m$--dimensional areas ($m < n$), sizes.  Suppose that
one term of the four--dimensional space--time vacuum
action is proportional to the volume of a fundamental element:
\be
\label{vol}
S_{vac} = A\, \hbar^{-1} G^{-2} c^6 \int\limits_{FE}  \sqrt{-\eta}\, d V,
\ee
where $A$ is a numerical factor.  Here $d V$ and
$\sqrt{-\eta}$ are continual space values. Integrating
in (\ref{vol}) is over one region from some covering
of the continual space--time. This term is analogous
to "space--time foam" action proportional to the
volume~\cite{hawking}.  But this term of an action is
unsufficient for description the equilibrium
configuration of the space--time of FFFE. Total actions
(\ref{vol}) of all configurations from the
set of coverings are equal. The action minimum must correspond to
the classical configuration, i.e. the continual
space--time configuration in the considered case.

The complete expression for the space--time action,
meeting this requirement, must contain other terms
besides the volume term. The possible term is the one
proportional to the total $n-1$--~dimensional area of
FFFE. In this supposition the vacuum space--time
action is
\be
S = \alpha \sum_i V_i + \beta \sum_i (S_{n-1})_i,
\ee
where $\sum_i$ is summarizing over all FFFEs.

Let us direct the way, on which the Nambu--Goto term
of the string action ~\cite{gsw} might be obtained from
the space--time action (\ref{vol}).  The expression of
the action of a space--time element for this analysis
is required. This action is the average value of an
action $S$ with the functional integrating
(\ref{func}) using. This action is denoted by
$S_{FFFE}$:
\be
S_{FFFE} = <S_{\{a\}}>
= A\,\hbar\,L_{pl}^{-4}
\int \limits_{FFFE} \sqrt{-\eta}\, dx^1 dx^2 dx^3 dx^4
\ee
for the four--dimensional space--time.

A string in the space--time of FFFEs can be considered
as an excitation of a number of FFFEs forming
one--dimensional space--like structure (in the meaning of
FFFE space--time).  In the own reference frame the
action of this excitation is represented in the form
\be
\label{SFFFE}
S = A\, \hbar\, L_{pl}^{-4} \int\limits_{FFFEs}
 \sqrt{-\eta}\, dx^1 dx^2 dl\, d\tau,
\ee
where $\tau$ is the own time, $l$ is the own
space--like coordinate of an excitation, $x^1, x^2$
are the transverse space--like coordinates.  Here
integrating is over a set of FFFEs, participated in
the excitation propagation. Integrations over
transverse coordinates yields the average values of FFFE sizes,
i.e. $l_f$. In supposition $l_f \simeq L_{pl}$
we might obtain ($\gamma$ is
the two--dimensional metric tensor determinant):
\be
\label{NG}
S = A\, \hbar\, L_{pl}^{-2} \int dl\, d\tau\, \sqrt{-\gamma},
\ee
i.e. the Nambu--Goto action for a string.
This result is not completely correct
due to the transformation problem of the
four--dimensional metric tensor determinant $\eta$ in
(\ref{SFFFE}) to the two--dimensional one $\gamma$ in
(\ref{NG}) and absence of the correct definition of
one--dimensional integrating.  In this concept the
$p$--branes are considered as the $p$--dimensional
space--like excitations of FFFEs, and the volume term
(\ref{SFFFE}) of the FFFE space--time excitation yields
the bosonic term of $p$--branes action analogically.

Consideration of the strings as the excitations
of the quantized space--time is the step to the
understanding of the superstrings properties at the
Plank distances.  With this strings and
$p$--branes consideration all these objects are
identical at the Plank distances because the
excitation of one element does not have dimension in the
sense of FFFE space--time.

The author thanks M.S. Orlov,
A.V. Klochkov, E.V. Klochkova, A.V. Sokolov, G.S. Sokolova,
A.B. Vankov for their
friendly support during the work time, A.A. Amerikantsev,
M.E. Golod, A.N. Lobanov
for their technical assistance and M.A. Tyntarev for
useful discussions.

\end{document}